\documentclass[epj,referee]{svjour}
\usepackage{graphics}

\usepackage [dvips]{graphicx}
\usepackage{amsmath}
\usepackage{amssymb}
\usepackage{epsfig}

\addtocounter{secnumdepth}{1}
\begin{document}
\newcommand{\br}{\bar{r}}
\newcommand{\bbeta}{\bar{\beta}}
\newcommand{\bgamma}{\bar{\gamma}}
\newcommand{\bE}{{\bf{E}}}
\newcommand{\bR}{{\bf{R}}}
\newcommand{\bS}{{\bf{S}}}
\newcommand{\bT}{\mbox{\bf T}}
\newcommand{\bt}{\mbox{\bf t}}
\newcommand{\half}{\frac{1}{2}}
\newcommand{\summ}{\sum_{m=1}^n}
\newcommand{\sumqno}{\sum_{q\neq 0}}
\newcommand{\tsum}{\Sigma}
\newcommand{\bsA}{\mathbf{A}}
\newcommand{\bsV}{\mathbf{V}}
\newcommand{\bsE}{\mathbf{E}}
\newcommand{\bsT}{\mathbf{T}}
\newcommand{\bsZ}{\hat{\mathbf{Z}}}
\newcommand{\bse}{\mbox{\bf{1}}}
\newcommand{\bspsi}{\hat{\boldsymbol{\psi}}}
\newcommand{\cdottt}{\!\cdot\!}
\newcommand{\deltaR}{\delta\mspace{-1.5mu}R}

\newcommand{\bGamma}{\boldmath$\Gamma$\unboldmath}
\newcommand{\dd}{\mbox{d}}
\newcommand{\ee}{\mbox{e}}
\newcommand{\p}{\partial}

\newcommand{\Rav}{R_{\rm av}}
\newcommand{\Rc}{R_{\rm c}}

\newcommand{\la}{\langle}
\newcommand{\ra}{\rangle}
\newcommand{\rao}{\rangle\raisebox{-.5ex}{$\!{}_0$}}  
\newcommand{\rae}{\rangle\raisebox{-.5ex}{$\!{}_1$}}
\newcommand{\raG}{\rangle_{_{\!G}}}
\newcommand{\rainr}{\rangle_r^{\rm in}}
\newcommand{\beq}{\begin{equation}}
\newcommand{\eeq}{\end{equation}}
\newcommand{\bea}{\begin{eqnarray}}
\newcommand{\eea}{\end{eqnarray}}
\def\lsim{\:\raisebox{-0.5ex}{$\stackrel{\textstyle<}{\sim}$}\:}
\def\gsim{\:\raisebox{-0.5ex}{$\stackrel{\textstyle>}{\sim}$}\:}


%
\title{Statistical properties
of planar Voronoi tessellations}
\author{H.J. Hilhorst 
}                     
%
%
\institute{Laboratoire de Physique Th\'eorique, 
B\^atiment 210,
Univ Paris-Sud and CNRS,
91405 Orsay, France}
\date{Received: date / Revised version: date}
%
\abstract{
I present a concise review 
of advances realized over the past three years
on planar Poisson-Voronoi tessellations. These encompass 
new analytic results, a new Monte Carlo method, and 
application to experimental data.
} 
\maketitle

\section{Introduction} 
\label{secintroduction}
\vspace{5mm}

\begin{figure}
\resizebox{0.75\columnwidth}{!}{%
\includegraphics{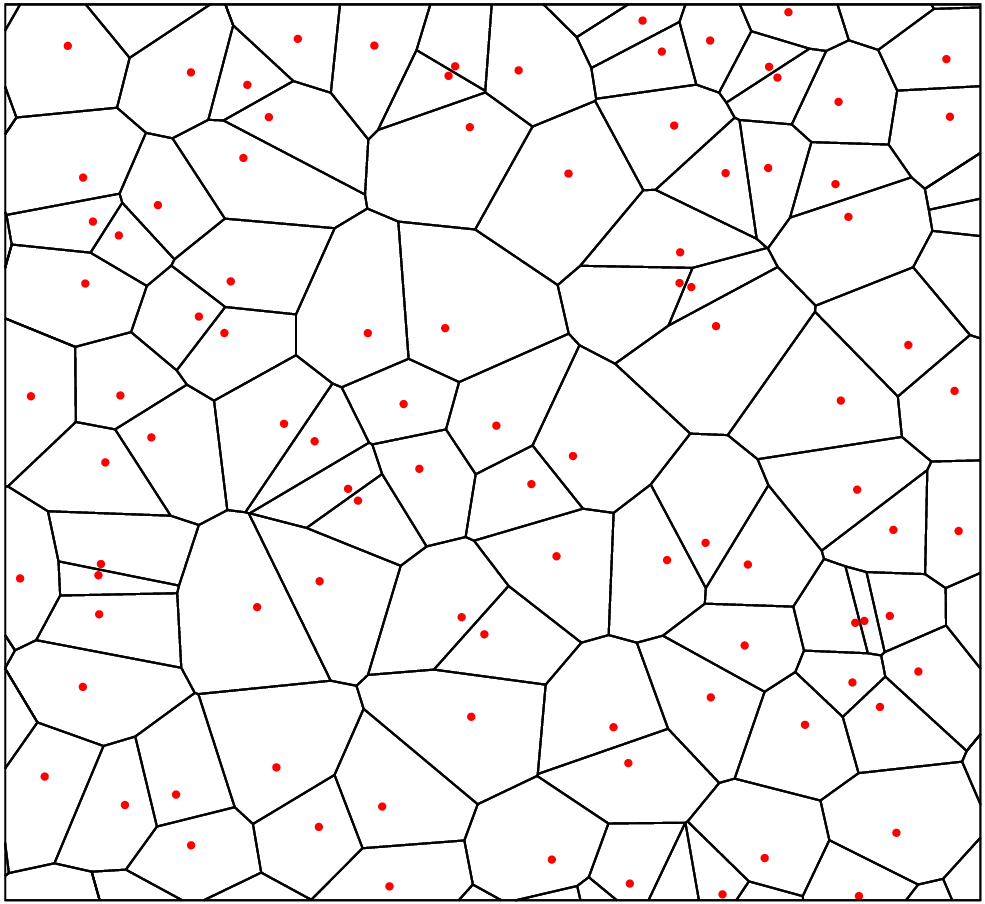}
}
\caption{Voronoi tessellation of a set of point particles.}
\label{figintrowh}       
\end{figure}

In this talk I will concisely review a coherent collection
of new results on planar Voronoi cells 
obtained over the last three years
\cite{Hilhorst05a,Hilhorst05b,Hilhorst06,Hilhorst07}.
In Fig.\,\ref{figintrowh} you see a {\it Voronoi tessellation}:
the set of dots is given and I will refer to them as
`centers' or
`point particles.' Perpendicular bisectors have been constructed to the line
segments connecting nearby particles. The bisectors meet in trivalent
vertices (unless there is accidental degeneracy) 
and partition the plane into convex polygons
called {\it Voronoi cells}. 

In physics Voronoi cells have, roughly, two broad classes of applications.
 (i) They may directly model cellular structures, whether
occurring naturally or synthesized; examples are biological tissue
or soap froths.
 (ii) They may serve as a tool of analysis.
For example, in a dense configuration of real physical particles
they define lattice defects; and if you decide
to study your favorite theory on a lattice
of randomly located spatial sites, the Voronoi construction is the natural
way to define nearest-neighbor relations between these sites. 
Longer lists of applications may be found in the literature
\cite{Okabeetal00,Rivier93,Hilhorst05b}.

When the point particles are distributed independently and uniformly
-- which I will henceforth stipulate --,
the resulting tessellation is called a {\it Poisson-}Voronoi tessellation.
Real systems will of course most often exhibit deviations from this simple
mathematical model:
soap froths evolve with time and their cells are not constructed around a
center; real particles have a minimum distance of approach; {\it etc.}
However, the Poisson-Voronoi tessellation, being the simplest one to study, 
is to cellular systems what the Ising model is to magnetism.

\begin{figure}
\resizebox{0.85\columnwidth}{!}{%
  \includegraphics{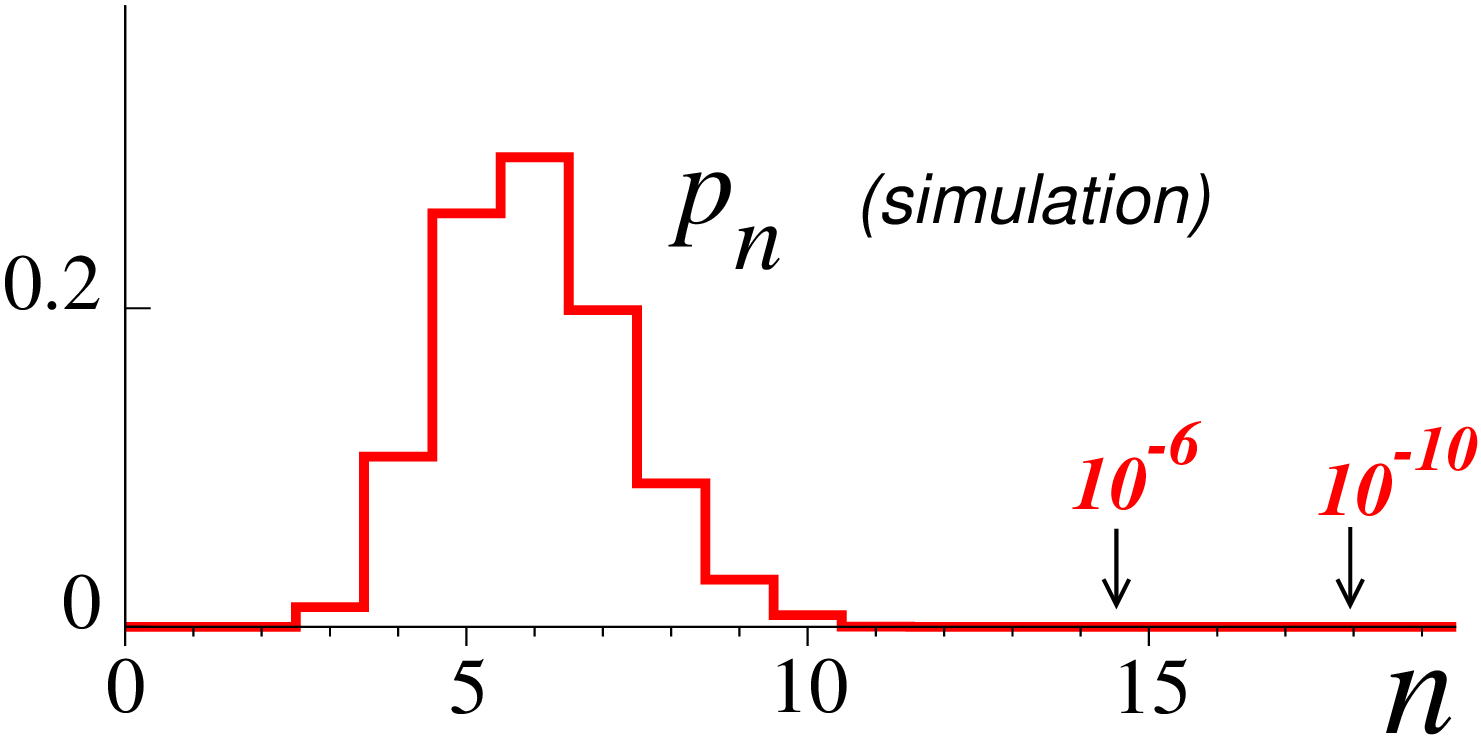}
}
\caption{Histogram of the sidedness probability $p_n$.}
\label{figpn5transp}       
\end{figure}

Over the last 50 years, starting with the work of Meij\-ering 
\cite{Meijering53},
many analytical properties of the Voronoi cell have been determined.
These include the statistics of the perimeter segments, of the angles at the
vertices, the cell area, and so on
(an overview is given in Ref.\,\cite{Okabeetal00}).
However, the Voronoi cell's most prominent property,
{\it viz.} its probability $p_n$ of being $n$-sided,
is extremely difficult to access analytically.
The basic expression for $p_n$ is a $2n$-dimensional integral on the 
positions\, $\vec{R}_m\,\, (m=1,\ldots,n)$ of 
$n$ point particles neighbor\-ing a central one
placed in the origin. 
A coupling between the $\vec{R}_m$ arises from the condition that the
resulting cell be $n$-sided. Hence the expression for
$p_n$ is on the same footing 
as a partition function of $n$ interacting particles in two dimensions.
No exact evaluation is known, but Monte Carlo simulations 
have led to the histogram of Fig.\,\ref{figpn5transp}. 
It shows that $p_n$ peaks at $n=6$ and decreases very rapidly for 
larger $n$. It is as small as $10^{-6}$ for $n\approx 15$ and
$10^{-10}$ for $n\approx 18$. Drouffe and Itzykson (DI)
\cite{DI84,ID89} devised a
Monte Carlo algorithm that yielded
$p_n$ with a single-digit
accuracy up to $n\approx 25$.
The asymptotic decay of $p_n$ for large $n$
has been a subject of speculations. Many authors fitted $p_n$ to a decaying
exponential or stretched exponential, often multiplied by
a power. DI proposed $p_n \sim n^{-\alpha n}$ with $\alpha\approx 2$.

\section{Mathematics}

A new line of research began with the mathematical challenge
of {\it obtaining the analytic asymptotic expression of $p_n$ 
for\, $n\to\infty$.} The calculation,
although involving only classical tools of analysis,
is of considerable complexity \cite{Hilhorst05a,Hilhorst05b}.
I present briefly its main ideas.

\begin{figure}
\resizebox{0.75\columnwidth}{!}{%
  \includegraphics{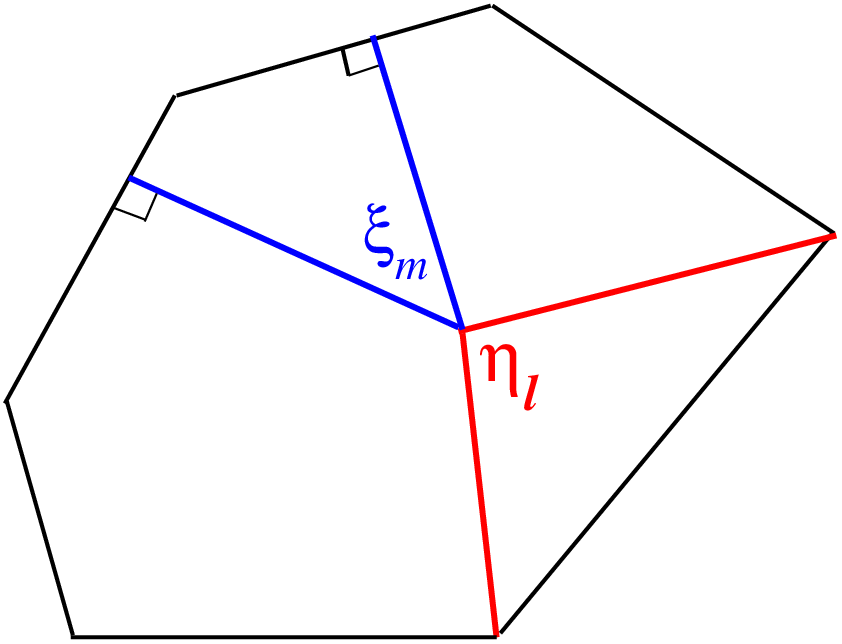}
}
\caption{The angles $\xi_m$ and $\eta_{\ell}$ (with $m,\ell=1,\ldots,n$)
that define the $n$-sided Voronoi cell up to a radial scale factor.}
\label{figxieta7}       
\end{figure}

A key role is played by the angles $\xi_m$ and $\eta_\ell$
defined in Fig.\,\ref{figxieta7}. They define the cell up to a
radial scale factor. 
The obvious sum rules 
$\sum_{m=1}^n\xi_m=2\pi$ and $\sum_{\ell=1}^n\eta_{\ell}=2\pi$ represent
infinitely weak constraints in the limit $n\to\infty$. 
It appears that in that limit, to zeroth order in $n^{-1}$,
the angles $\xi_m$ and $\eta_{\ell}$ become a set of
$2n$ independent variables, 
the distributions $P_1^{(0)}(n\xi_m/2\pi)$ and $P_2^{(0)}(n\eta_\ell/2\pi)$ 
being given by
\beq
P_1^{(0)}(x) = 4x\,\ee^{-2x}, \qquad
P_2^{(0)}(y) = \ee^{-y}, \qquad x,y>0.
\label{defP0}
\eeq
This independence is nontrivial: 
the $\xi_m$ and $\eta_{\ell}$ overlap, so how can
they be uncorrelated? This property 
is best initially introduced as a hypothesis,
to be confirmed self-consistently later by the solution procedure.

The independence of the angles 
can be shown to imply that
as $n\to\infty$ the shape of the $n$-sided cell
approaches a circle  with probability $1$. 
From an analysis of the radial integral
in the expression for $p_n$ one concludes that 
(for point particle density $\lambda$) this circle is of
the radius $R_{\rm c}(n)=\sqrt{n/4\pi\lambda}$.

After writing $p_n$ in terms of the variables of
integration $\xi_m$ and $\eta_\ell$, one can express it in the factorized
form 
\beq
p_n=C_np_n^{(0)}, \qquad p_n^{(0)}=\frac{ 2(8\pi^2)^{n-1} }{ (2n)!},
\label{split}
\eeq
where $p_n^{(0)}$ results from integration with respect 
to the $P_1^{(0)}$ and $P_2^{(0)}$.
Hence the problem of finding $p_n$ has been replaced with 
that of finding $C_n$. This problem involves the ${\cal O}(n^{-1/2})$
and ${\cal O}(n^{-1})$ corrections to the infinite-$n$ behavior.
It appears that for $C_n$ one can set up a perturbation expansion which
shows that 
\beq
C_n = C[\,1\,+\,{\cal O}(n^{-1})], \qquad n\to\infty,
\label{Cnaspt}
\eeq
where $C$ can be expressed elegantly as
\beq
C=\prod_{q=1}^\infty \Big(1-\frac{1}{q^2}+\frac{4}{q^4}\Big)^{-1}
 =0.344\,347\ldots
\label{resultC}
\eeq
This infinite product is in fact the `elastic' 
configurational partition function of the cell perimeter. 
The factor of index $q$ in (\ref{resultC})
stems from deviations from circularity having a wavelength $2\pi R_{\rm c}/q$ 
along the perimeter.

There is an interesting corollary. It says
that there is a continuum limit in which the cell perimeter, expressed as a
function $R (\phi)$ of the polar angle $\phi$, satisfies 
\beq
\frac{\dd^2 R(\phi)}{\dd\phi^2} = L(\phi),
\label{nbbg}
\eeq
where $L(\phi)$ is Gaussian noise which is
white to order zero but has colored order $n^{-1}$ corrections.
The solution of (\ref{nbbg}), under appropriate conditions 
pertaining to the average of $R(\phi)$ 
and to its periodicity \cite{Hilhorst05b}, 
produces the full functional probability distribution
${\cal P}[R(\phi)]$ of an arbitrary perimeter $R(\phi)$.
Hence the asymptotic determination of $p_n$
leads to a complete understanding
of the behavior of the large $n$-sided cell.

Finally, the mathematical methods employed here
are applicable to other problems that have arisen in mathematics and in
physics. I mention two of them.

{\it The Crofton problem.} Let a plane be traversed by intersecting
straight lines, distributed randomly and uniformly. 
This is another way of partitioning it into convex cells.
The `Crof\-ton problem' is the question of determining the probability 
$p_n^{\rm Cr}$ for the cell containing the origin to be $n$-sided.
This was done in Ref.\,\cite{CalkaHilhorst07}, 
again in the limit of large $n$. 

{\it The problem of extremal points.}
Let $n$ points be distributed randomly and uniformly in the unit disk.
Then can one determine the probability $p_n^{\rm ext}$ that 
the convex envelope of this set is an $n$-sided polygon?
Work on this problem is in progress.

\section{Monte Carlo method}

Attempts to estimate $p_n$ by Monte Carlo simulation
have been numerous (see Ref.\,\cite{Okabeetal00}).
Straightforward methods star\-ting from a random set of centers
suffer from the fact that many-sided cells are very rare. 
More sophisticated methods
like those of Ref.\,\cite{DI84} aim at
directly generating $n$-sided cells for an $n$ fixed in advance.
   
The derivation of the asymptotic result 
(\ref{Cnaspt})-(\ref{resultC}) for $C_n$
has as its starting point a {\it non}\,asymptotic expression of the form
\beq
C_n=\langle W_n \Theta_n \rangle,
\eeq
valid for all $n=3,4,\ldots$. Here $\langle\ldots\rangle$ is an average with
respect to the $P_1^{(0)}$ and $P_2^{(0)}$. Furthermore
$W_n$ is a weight that can be expressed \cite{Hilhorst07}, 
through a series of equations, in
terms of the sets of angles $\{\xi_m\}$  and $\{\eta_\ell\}$,  
and $\Theta_n$ is a projector: $\Theta_n=1$ 
if a certain condition
on the angles is satisfied, and $\Theta_n=0$ if not. This condition 
requires the $n$ points $\vec{R}_m$ all to be located such
that they contribute a
nonzero segment to the perimeter.
All previous `fixed $n$'
Monte Carlo methods described in the literature
run into such an acceptance criterion. 
In all proposed cases, the projector rejects an ever
larger fraction of configurations as $n$ gets larger: this is the 
phenomenon of {\it attrition}, well-known (and deplored) in many 
Monte Carlo studies.

The particular split (\ref{split})
that I made into a zeroth  order problem and a remainder
now appears to be the right one \cite{Hilhorst07}. 
If the $\xi_m$ and $\eta_\ell$ are chosen 
from the zeroth order distributions (\ref{defP0}), then
{\it the problem of
attrition in the large-$n$ limit is eliminated\,}: 
as $n$ grows, the fraction of accepted configurations tends to
unity! In practice, less than $1\%$ of the configurations is
rejected for $n \geq 20$, and less than $0.01\%$ for $n\geq 40$.  

Two things then become possible.

(i) To accurately determine the sidedness probabilities $p_n$ for arbitrarily
large $n$. Tables with values of $p_n$ to at least four decimal places are
given in Ref.\,\cite{Hilhorst07}  in the range $3\leq n\leq 1600$.
For high $n$ these probabilities become unphysically small:
one has $p_{100}=5.269\times 10^{-188}$ and 
$p_{1000}= 6.36 \times 10^{-3841}$ ({\it sic!\,}),
but I will show below which rewards can be gained from studying them.

(ii) To Monte Carlo generate
typical $n$-sided cells for arbitrary {\it a priori} given $n$,
together with their `natural' environment of other cells.
This is done as follows. Angles $\xi_m$ and $\eta_\ell$ are randomly
picked from the zeroth order distribution (but subject to the sum rules).
If they pass the acceptance criterion, the cell perimeter is
constructed. By radial scaling
the cell radius is given its most probable value $R_{\rm c}$. 
The positions of the central point particle and its
first neighbors are now fixed. Next, 
all other point particles, in order that they do not interfere
with the central Voronoi cell, are
excluded from the `fundamental domain' associated with that cell, 
{\it i.e.} from the
union of the $n$ disks centered at the cell vertices and passing
through the origin; however, those other point particles
(which are second and higher order neighbors to the central one),
occupy the remaining region of the plane uniformly and randomly.
Hence, after Monte Carlo generation of
their positions, the construction of the full
Voronoi tessellation may be completed by any of the existing algorithms.

\begin{figure}
\resizebox{0.85\columnwidth}{!}{%
  \includegraphics{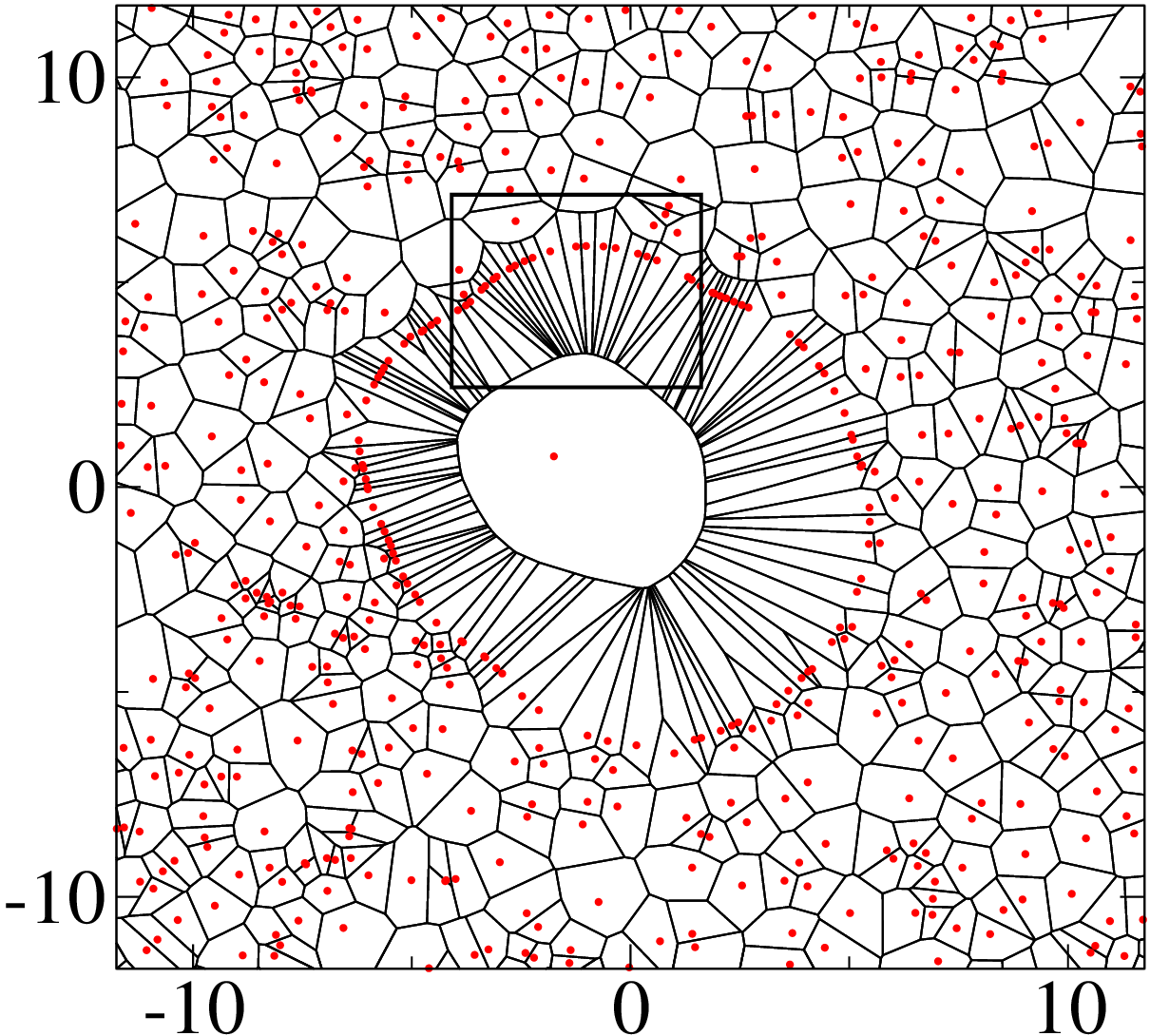}
}
\caption{Snapshot of a $96$-sided Voronoi cell.}
\label{figureMC7}      
\end{figure}

Fig.\,\ref{figureMC7} shows a $96$-sided Voronoi cell constructed in this
way. Snapshots of cells
with $n=24, 48, 96$, and $1536$ may be found in Ref.\,\cite{Hilhorst07}.

\section {Aboav's law}

Aboav's law, formulated in 1970 \cite{Aboav70}, holds that 
the neighbor of an $n$-sided cell has itself an average number $m_n$
of sides given by
\beq
m_n=a+bn^{-1},
\label{Aboavslaw}
\eeq
where $a$ and $b$ are positive constants. 
Eq.\,(\ref{Aboavslaw}) used to be, 
and is still often, formulated as $nm_n=an+b$, and is therefore
called the `linear law'.
Since $m_n$ decreases with $n$, it
says that {\it ``many-sided cells tend to have
few-sided neighbors, and conversely.''}
Two-cell correlations have been measured in a large variety of
experimental systems and
found in agreement with Aboav's law for parameters  
in the range $4.5\lesssim a\lesssim 5.3$ 
(in this context $5$ has been called a `magic number') 
and $5.7\lesssim b\lesssim 8.5$. 

The first one to present heuristic arguments for the validity of
(\ref{Aboavslaw}) was 
Weaire \cite{Weaire74}, whence the alternative name `Aboav-Weaire law.'
Other `proofs' of this law (see Ref.\,\cite{Hilhorst06} for some references) 
appeal to mean field approximations 
or proceed by maximizing a hypothesized entropy functional. 
While some workers consider Aboav's law as no more than a
linear approximation to some unknown curve, others have attributed to it
a more fundamental status, as is clear from numerous statements in the
literature: {\it ``In this paper we present an alternative derivation of the
  Aboav-Weaire law''} \cite{EdwardsPithia94}, or
{\it ``In all known natural and man-made structures, it is found empirically
  that $nm_n$ is linearly related to $n$}'' \cite{Dubertretetal98}, and many
others. 

\begin{figure}
\resizebox{0.85\columnwidth}{!}{%
  \includegraphics{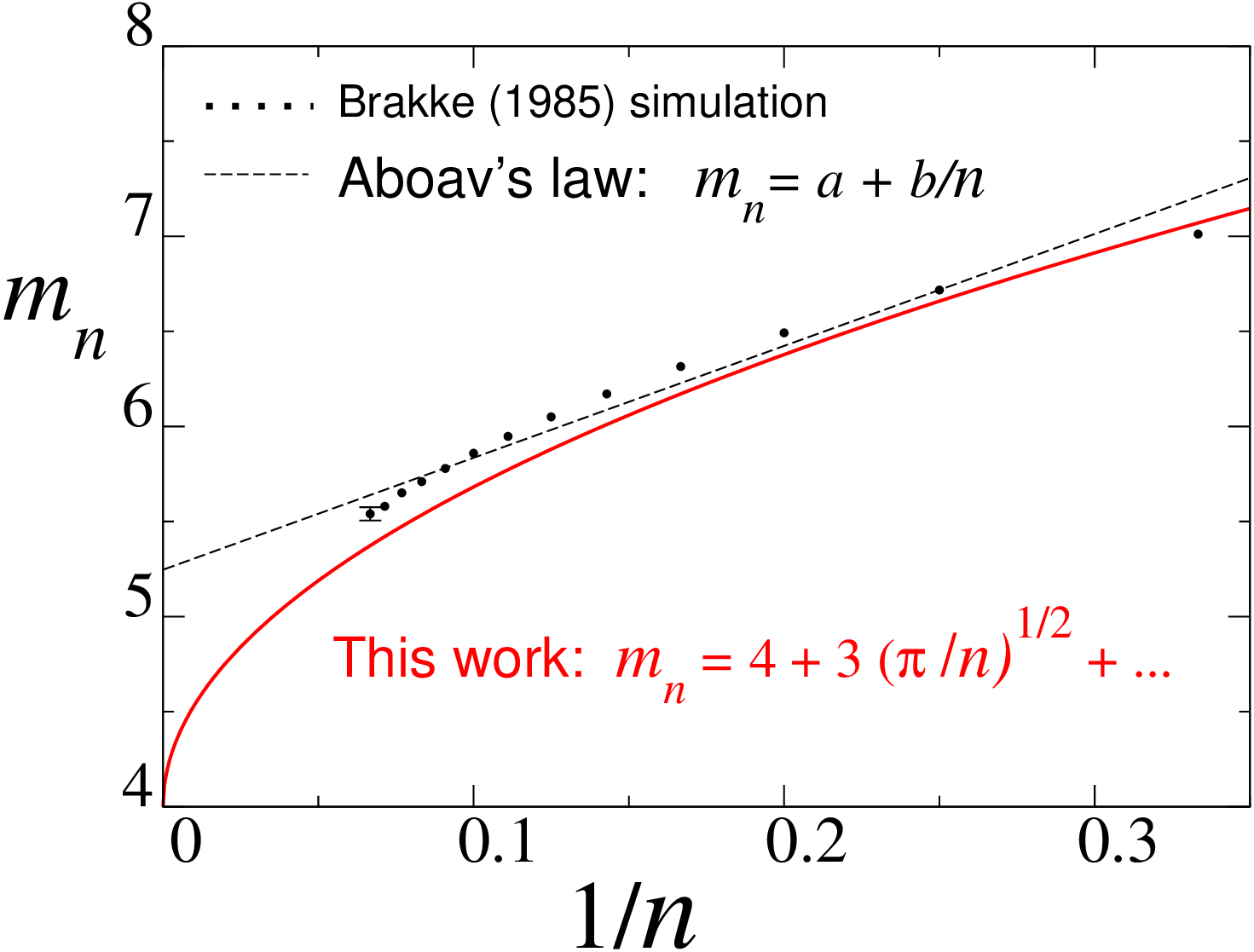}
}
\caption{Monte Carlo data, Aboav's law, and this work's asymptotic expression
(\ref{resultmn}) for $m_n$.}
\label{figbah5}   
\end{figure}

Now it has been known for over twenty 
years that Aboav's law does {\it not\,} hold
exactly for the Poisson-Voronoi tessellation. 
This is borne out by Fig.\,{\ref{figbah5}}:
numerical simulation shows
unambiguously that the $m_n$ curve has a very small but distinct downward
curvature. It is so small that, if also present in nature, 
most experimental data would not detect it.

In face of this, the defenders of Aboav's law as a basic truth hold that
the Poisson-Voronoi tessellation, because of being constructed around
random centers, does not
correspond to any true physical structure. In real cellular structures, so they
argue, any initial randomness has always undergone a relaxation process
(for example to relieve internal stress); and this then would have induced the
validity of Aboav's law.
\begin{figure}
\resizebox{0.65\columnwidth}{!}{%
  \includegraphics{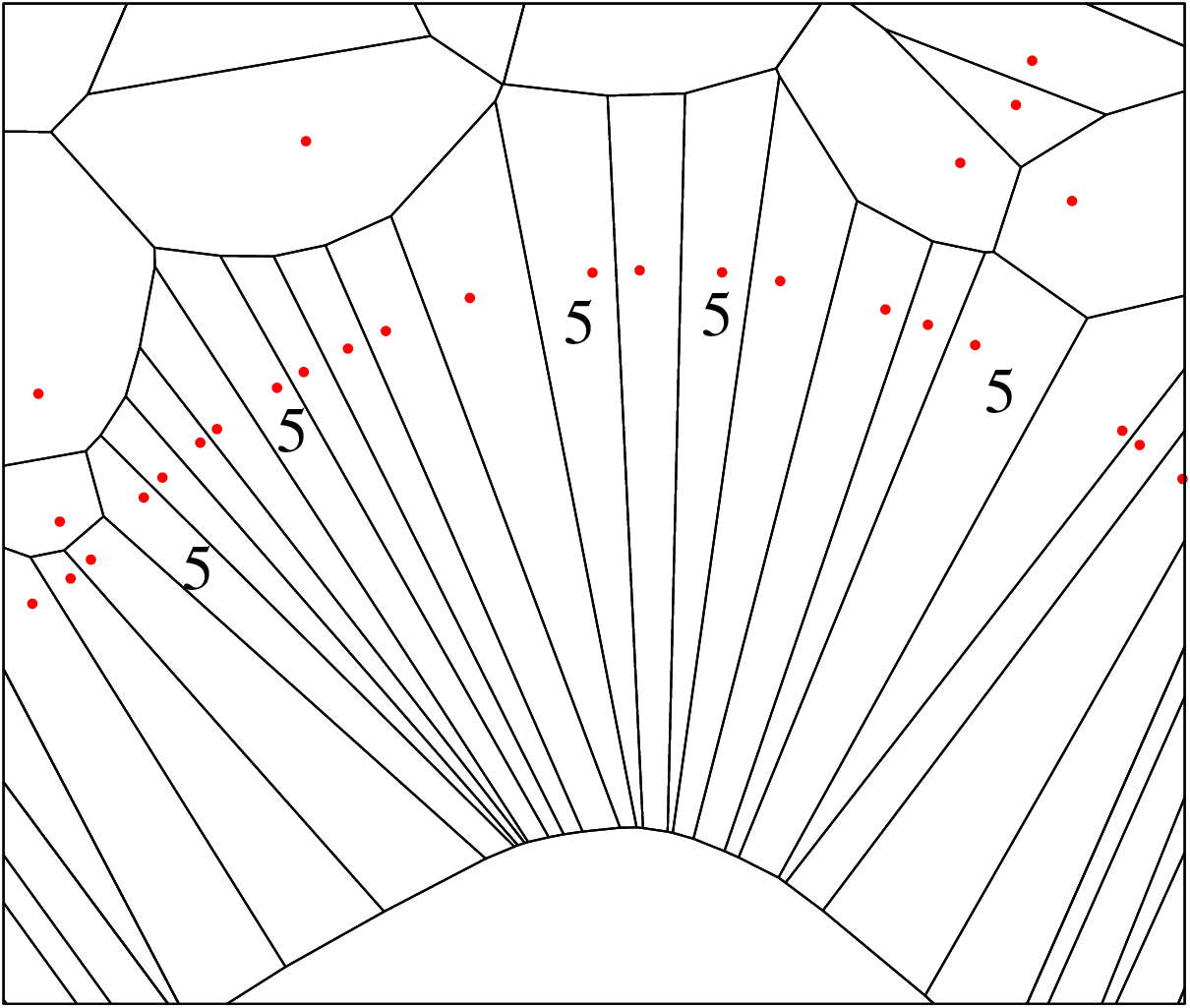}
}
\caption{Zoom onto the box in Fig.\,\ref{figureMC7} showing some
  first-neighbor cells to the central cell.}
\label{figureMC8a} 
\end{figure}

\subsection{Aboav's law and the Poisson-Voronoi tessellation}

Before returning to real physical systems, 
let us first see what the present theory, further developed in 
Ref.\,\cite{Hilhorst06}, has to say about the validity of Aboav's law
for the Poisson-Voronoi tessellation. The picture of Fig.\,\ref{figureMC8a}
is useful for the argument.
It shows a phenomenon that becomes ever more pronounced
as $n$ gets larger (whence the need to consider very large $n$;
see \cite{Hilhorst07} for a picture with $n=1536$):
the first neighbors of the central cell become more and more
elongated while their width goes down to order $1/\sqrt{n}$. 
Since the second neighbors of the central cell have 
dimensions of order unity (for unit particle density),
the geometry dictates that most first neighbors must become four-sided.
The fraction of them that is five-sided is only of order $1/\sqrt{n}$;
this is because
five-sidedness occurs only when 
a first-order neighbor is adjacent to two
second-order neighbors. In Fig.\,\ref{figureMC8a}
all five-sided cells have been marked.

It follows that $m_n$ is equal to $4$ plus ${\cal O}(n^{-1/2})$ corrections.
The latter require a nontrivial calculation \cite{Hilhorst06} and 
one finds
\beq
m_n=4+3\sqrt{ \pi/n } +\ldots, \qquad n\to\infty.
\label{resultmn}
\eeq
In Fig.\,\ref{figbah5} the Monte Carlo data for $m_n$
(due to Brakke \cite{Brakke}) are compared to 
the first two terms of the asymptotic expansion (\ref{resultmn}), 
as well as to the best
linear two-parameter fit of type (\ref{Aboavslaw}).
Clearly the asymptotic theory that I developed
explains for the first time why $m_n$ must be
curved; moreover, it predicts correctly the order of magnitude of this 
curvature \cite{Hilhorst06}.

\subsection{Aboav's law and experimental systems}

Under suitable conditions, polystyrene latex 
spheres of diameter
$\approx 1 \mu m$, when trapped at a water/air interface,
will on a time scale of hours undergo a process of
diffusion limited colloidal aggregation (DLCA). 
The experiment was performed by Earnshaw and coworkers
\cite{EarnshawRobinson94,EarnshawRobinson95,EHR96} in the 1990's 
and was simulated very recently by Fern\'andez-Toledano {\it et al.}
\cite{FTetal05}.
Snapshots of the system were taken 
at regular time intervals and 
the center of gravity of each cluster of particles was determined. 
Then the Voronoi cells belonging to this set of centers were
constructed and $m_n$ was obtained.
\begin{figure}
\resizebox{0.85\columnwidth}{!}{%
  \includegraphics{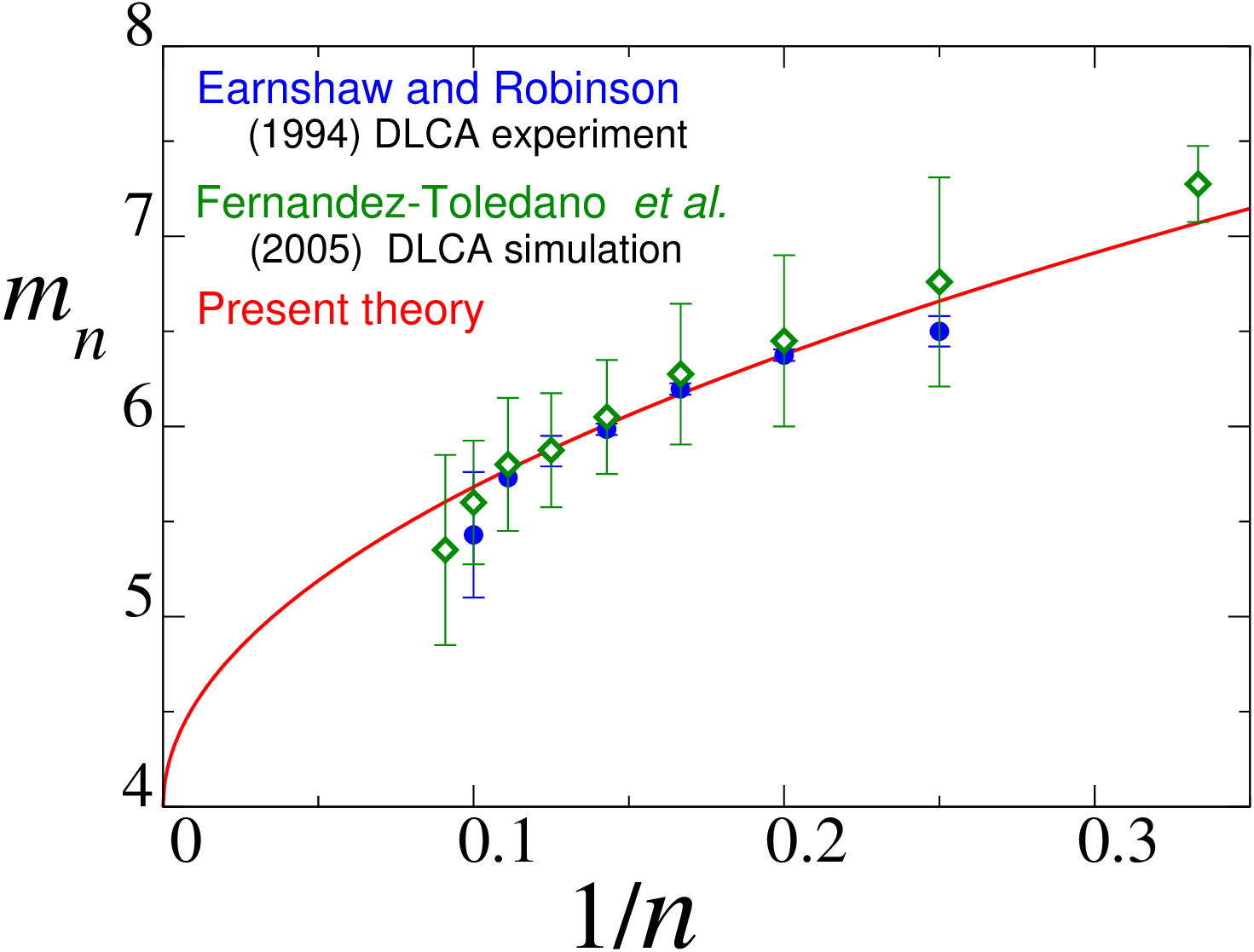}
}
\caption{Experimental and simulational $m_n$ data compared to the asymptotic
Poisson-Voronoi curve (\ref{resultmn}).}
\label{figefh2}
\end{figure}
Fig.\,\ref{figefh2} shows the experimental DLCA
data (dots) of Ref.\,\cite{EarnshawRobinson94}
taken after 60 seconds, as well as the corresponding simulation data 
(diamonds) of Ref.\,\cite{FTetal05}. Also shown 
is the same theoretical curve (solid line) as in Fig.\,\ref{figbah5}.
In view of the experimental and numerical error bars,
this figure does not {\it prove\,} that the DLCA system is of the
Poisson-Voronoi type.
I believe however that, as an explanation for these data,
the present zero-parameter curvature-predicting theory 
is preferable to the linear two-parameter fit (\ref{Aboavslaw}).

A final point is worth discussion.
In the experiment, the clusters whose centers of gravity
serve as the `point particles' of the Voronoi construction,
have some finite diameter $d$. Is this a problem?
The experiment is done, typically, at an area coverage of about $10\%$
\cite{EarnshawRobinson94}, and hence
at a cluster number density $\lambda$ such that 
$\pi d^2\lambda/4 \approx 0.1$.
In the Poisson-Voronoi 
model, for a point particle density $\lambda$, the typical distance
between two first neighbors of an $n$-sided cell
is $\ell_n \equiv 2\pi(2R_{\rm c})/n=\sqrt{4\pi/n\lambda}$.
The results of the present asymptotic theory can be expected to be
applicable only
for $n \lesssim n^*$, where $n^*$ is a sidedness such that 
$\ell_{n^*} \approx d$. The point particle/cluster
density $\lambda$ drops out of this relation and we find $n^*\sim 100$.
Taking into account the possibly fractal structure of the clusters
may lower this estimate of $n^*$, 
but the $n$ values 
of Fig.\,\ref{figefh2} are less than $n^*$ by what seems a safe margin.
Crossover between the ideal Poisson-Voronoi behavior and
a finite-diameter theory was briefly discussed in Ref.\,\cite{Hilhorst06}.

\section{Conclusion}

What we learn from this is that in other experimental situations, too,
as the range and
the precision of the data increase, {\it a curved $m_n$ law\,}
may well be expected and must indeed be looked for.

And, on a more general level, that provided you ask the right questions, 
{\it the patterns induced by random points in a plane
are challenging!}


\begin{thebibliography}{}

\bibitem{Hilhorst05a}
H.J.~Hilhorst,
{\it J. Stat. Mech.} (2005) L02003.

\bibitem{Hilhorst05b}
H.J.~Hilhorst,
{\it J. Stat. Mech.} (2005) P09005.

\bibitem{Hilhorst06}
H.J.~Hilhorst, {\it J.~Phys.~A\,} {\bf 39} (2006) 7227.

\bibitem{Hilhorst07}
H.J.~Hilhorst,
{\it J. Phys. A\,} {\bf 40} (2007) 2615.

\bibitem{Okabeetal00}
A.~Okabe, B.~Boots, K.~Sugihara, and S.N.~Chiu,
{\it Spatial tessellations: concepts and applications of
Voronoi diagrams,} second edition (John Wiley \& Sons Ltd., 
Cichester, 2000). 

\bibitem{Rivier93}
N. Rivier in: {\it Disorder and Granular Media,}
eds. D. Bideau and A. Hansen, Elsevier (1993).

\bibitem{Meijering53}
J.L. Meijering, 
{\it Philips Research Reports\,} {\bf 8} (1953) 270.

\bibitem{DI84}
J.M.~Drouffe and C.~Itzykson,
{\it Nuclear Physics\,} {\bf B235 [FS11]} (1984) 45.

\bibitem{ID89}
C.~Itzykson and J.M.~Drouffe,
{\it Statistical field theory} (Cambridge University Press,
Cambridge, 1989), Vol.\,2, ch.\,11.

\bibitem{CalkaHilhorst07}
P. Calka and H.J. Hilhorst, {\it in prepration.}

\bibitem{Aboav70}
D.A.~Aboav,
{\it Metallography\,} {\bf 3} (1970) 383.

\bibitem{Weaire74}
D. Weaire,
{\it Metallography\,} {\bf 7} (1974) 157.

\bibitem{EdwardsPithia94}
S.F. Edwards and K.D. Pithia,
{\it Physica A\,} {\bf 205} (1994) 577.

\bibitem{Dubertretetal98}
B. Dubertret, N. Rivier, and M.A. Peshkin,
{\it J. Phys. A\,} {\bf 31} (1998) 879.

\bibitem{Brakke}
K.A. Brakke (1986) {\it unpublished.}
Available on 
{\tt http://www.susqu.edu./brakke/aux/downloads/200.pdf}.

\bibitem{EarnshawRobinson94}
J.C. Earnshaw and D.J. Robinson,
{\it Phys. Rev. Lett.} {\bf 72} (1994) 3682.

\bibitem{EarnshawRobinson95}
J.C. Earnshaw and D.J. Robinson,
{\it Physica A\,} {\bf 214} (1995) 23.

\bibitem{EHR96}
J.C.~Earnshaw, M.B.J.~Harrison, and D.J.~Robinson,
{\it Phys. Rev. E\,} {\bf 53} (1996) 6155.

\bibitem{FTetal05}
J.C.~Fern\'andez-Toledano, A.~Moncha-Jord\'a, F.~Mart\'{\i}nez-L\'opez,
A.E.~Gonz\'alez, and R.~Hidalgo-\'Alvarez,
{\it Phys. Rev. E\,} {\bf 71} (2005) 041401.


\end{thebibliography}
\end{document}